# Dynamic Bidding Strategies with Multivariate Feedback Control for Multiple Goals in Display Advertising

Michael Tashman, Jiayi Xie, John Hoffman, Lee Winikor, Rouzbeh Gerami
Copilot AI
Xaxis
New York, USA
{michael.tashman, jiayi.xie, john.hoffman, lee.winikor, rouzbeh.gerami}@xaxis.com

## ABSTRACT

Real-Time Bidding (RTB) display advertising is a method for purchasing display advertising inventory in auctions that occur within milliseconds. The performance of RTB campaigns is generally measured with a series of Key Performance Indicators (KPIs)—measurements used to ensure that the campaign is cost-effective and that it is purchasing valuable inventory. While an RTB campaign should ideally meet all KPIs, simultaneous improvement tends to be very challenging, as an improvement to any one KPI risks a detrimental effect toward the others. Here we present an approach to simultaneously controlling multiple KPIs with a PID-based feedback-control system. This method generates a control score for each KPI, based on both the output of a PID controller module and a metric that quantifies the importance of each KPI for internal business needs. On regular intervals, this algorithm—Sequential Control—will choose the KPI with the greatest overall need for improvement. In this way, our algorithm is able to continually seek the greatest marginal improvements to its current state. Multiple methods of control can be associated with each KPI, and can be triggered either simultaneously or chosen stochastically, in order to avoid local optima. In both offline ad bidding simulations and testing on live traffic, our methods proved to be effective in simultaneously controlling multiple KPIs, and bringing them toward their respective goals.

## 1 Introduction

Real-Time Bidding (RTB) is a popular method for selling online display advertising inventory, typified by high-speed auctions that occur when an end-user requests a web page or element on their browser. Traditional advertising may rely on contracts, whereby an advertiser rents space over a fixed time period for an agreed-upon price. Some alternative methods, such as sponsored search and contextual advertising, enable advertisers to set a bid price for certain keywords or context elements. In this setup the ad space is given to the advertiser who has the highest bid associated with the relevant keywords or elements. RTB, by contrast, does not use pre-saved bids—instead, a bid request is made by a website at the time that a user attempts to load a web page. This request is sent from a Supply-Side Platform (SSP) through an ad exchange, which solicits bids from Demand-Side Platforms (DSPs), while providing impression-level features—such as country, time zone, domain name, and placement on page—and potentially some demographic information. Each advertiser's account on the DSP includes proprietary logic about the amount to bid given these impression-level features. In order for the winning ad to be delivered before the user's page finishes loading, ad auctions generally take place within 100ms [18]. Most auctions are second-price, directing advertisers to bid the value that they believe to be their true value of an impression [10].

The simple goal for much of display advertising is to receive a click. Advertisers may also seek conversion events—for example, a user may click an ad and then order a product from the resulting page; this is known as a post-click conversion. Conversion events are also possible without a click: a user who simply views an ad (and subsequently receives a cookie), and then goes on to buy the advertised product some time later, will be said to have completed a post-view conversion [13].

For a typical display advertising campaign, the probability of any given ad receiving a click or conversion is typically between 0.01% and 0.1% [13, 18]. Therefore, campaigns seek to purchase large quantities of display ads, and then use aggregate metrics to assess their overall cost and performance. Ads are priced in units of 1000—a unit known as CPM, or cost-per-mille. Performance is commonly measured in cost per click (CPC) or cost per action (CPA)—simply the amount spent divided by the number of clicks or conversion events, respectively [13]. Advertisers may seek a certain aggregate viewability—a metric defined as the percentage of impressions in which least 50% of pixels are visible on the user's display for at least 1 second [14]. Advertisers will likely also seek to ensure that all of the ad budget is spent by the planned end of the campaign, and that the budget is spent relatively smoothly.

Advertisers have several methods for improving KPIs. For example, in order to lower CPC or CPA they may apply a multiplier of less than 1 to all bids. In order to improve viewability, advertisers may set a viewability threshold: that is, decline to bid on any impression if the predicted probability that the ad will be viewable is below a certain percentage. Conversely, to ensure smooth pacing, advertisers may relax the aforementioned requirements to widen the pool of available inventory.

However, an attempt to improve one KPI may harm others: for example, reducing the bid multiplier to improve CPC may reduce addressable inventory and harm smooth pacing. Raising the viewability threshold may harm both pacing and CPC. If no solution is available that can bring all KPIs to goal, then it is essential to understand which KPIs are most important for the business. With this understanding, we may pursue the optimal attainable solution.

In order to address these challenges, we employ feedback control theory to generate dynamic bidding strategies—adapting to the non-stationary RTB environment in real time. Additionally, we present two methods to sequentially select which KPI to improve at a given time: Simple Sequential and Smart Sequential.

Our feedback-control method seeks the optimal attainable solution in an RTB environment through: (i) Using a PID controller to ascertain an error value for each KPI; (ii) acquiring from our internal trading team an objective ranking of each KPI's value to the business, and combining those to generate an overall goodness score for each KPI; (iii) On regular intervals, using the KPI "goodness" score to choose the KPI most in need of adjustment; (iv) Using a lookup table which, for each KPI, provides a set of rules that describe the direction to move the available levers in order to improve that KPI; (v) Using the PID error value as a control signal to input into an actuator module—where the actuator makes adjustments to some or all of the available levers in order to improve one or more KPI's, and where the magnitude and direction of the lever change(s) are directly derived from the control signal. The resulting feedback-control system is effective at bringing the most important KPIs to their goal, while adapting to unfamiliar market circumstances.

## 2  Background

In this section, we will discuss previous work relevant to this research, in the topics of feedback control of dynamic systems, and the sequential selection of KPIs to control.

### 2.1  Previous Works

Feedback control is a broad topic with a wide variety of applications, based on the goal of controlling a dynamic system through the repeated process of taking actions, observing feedback, and accounting for outside noise; its diverse uses include fields such as transportation, manufacturing, and robotics [1]. Zhang, Rong, et al. [2016] describes a method for using feedback control to improve KPIs in real-time display advertising. Their approach is a straightforward feedback loop, including a *controller* module, which assesses how close the system is to a goal, along with how quickly the observations are changing, and an *actuator* module, which uses the output of the controller as a basis for making adjustments to the system. Their approach validates that this method is effective in improving CPC in a real-time bidding environment.

Methods of joint optimization of multiple goals in advertising have been proposed in prior research, such as Geyik, et al. [2016] and Kitts, et al. [2017]. These approaches seek to improve multiple metrics by optimizing one at a time—our method, described in Section 4, functions in the same way. In Geyik, authors discuss a method to optimize pacing and viewability by regularly assessing whether delivery is acceptable, and if so, raising the dynamic filtering threshold for the viewability rate. This value, $\varphi(v)$, which we refer to as the *viewability threshold*, is a cutoff point: if the predicted probability that an impression will be viewable, $p(v)$, is less than $\varphi(v)$, the algorithm does not place a bid. Assuming that probability-of-view predictions are accurate, it follows that raising $\varphi(v)$ will induce an increase in aggregate viewability. It also follows that raising the viewability threshold will exclude a non-negative quantity of otherwise available inventory, and that at the maximum of $\varphi(v) = 1.0$, this will exclude all inventory. Therefore we have a viewability threshold, $\varphi(v)^*$, where $0 \leq \varphi(v)^* \leq 1$, such that any viewability threshold greater than $\varphi(v)^*$ will yield unacceptably low pacing. Since we can assume that, ceteris paribus, the highest possible aggregate viewability is most desirable, then it follows that we can optimize viewability and pacing if we have $\varphi(v) = \varphi(v)^*$. Geyik seeks to achieve this through a feedback-control method: their approach measures delivery on regular intervals; at each interval, raises $\varphi(v)$ by a small value ($\Delta$) when delivery is acceptable, or reduces $\varphi(v)$ by $\Delta$ when delivery becomes unacceptable. This should eventually result in $\varphi(v)$ approximating $\varphi(v)^*$.

Another approach, Kitts, et al. [2017], seeks to improve two or more KPIs simultaneously, in a way that will result in the greatest possible gain across all KPIs. Their basic reasoning is that satisfying KPIs is not really the end goal of the advertiser—instead, KPIs are a heuristic by which the advertiser may judge whether the campaign is targeting the kind of traffic that they believe is likely to be economically valuable. Therefore, KPIs should not be seen as a series of immutable constraints, but rather a multivariate optimization problem, in which the objective is to minimize overall error. Thus in a case with no feasible solution to all of the advertiser's required KPIs, the best approach may be to work to achieve good performance, while minimizing across-the-board KPI error as much as possible. Authors propose a method of calculating each KPI's error, with a step discontinuity to significantly de-weigh the error when the KPI is above goal rather than below. Then at every step, the system can work to improve the KPI most in need. This should ultimately converge to the optimal solution.

## 3  Feedback Control

Our goal is to help advertisers maximize the total value of winning impressions and meet certain key performance indicator (KPI) constraints, such as pacing, performance, and viewability. The pacing constraint requires that the budget is spent smoothly, and is fully spent at the end of a billing cycle. A performance constraint, such as CPC or CPA, casts a restriction on the cost of media relative to the target action. Given that advertisers can calculate the value to them of a target action, the use of performance goals ensures that an ad campaign generates sufficient value for the advertiser. Numerous performance KPIs exist; in order to focus the discussion without loss of generality, this paper focuses on only CPC and CPA, which are commonly used. The final constraint, viewability, casts another restriction on the quality of bought impressions: it requires that a reasonable fraction of bought impressions are actually viewable on end users' displays.

The problems can be formulated in terms of the following equations:

$$\max \sum_{i=1}^{N} x_i v_i,$$

$$s.t.\ spend = \sum_{i=1}^{N} x_i cost_i \leq B,$$

$$CPC = \frac{\sum_{i=1}^{N} x_i cost_i}{\sum_{i=1}^{N} x_i click_i} \leq C,$$

$$viewability = \frac{\sum_{i=1}^{N} x_i viewable_i}{\sum_{i=1}^{N} x_i} \geq V$$

Where: $x_i$ is an indicator variable, such that $x_i = 1$ when the advertiser wins the $i$-th ad opportunity, and $x_i = 0$ otherwise; $v_i$ indicates the value of the $i$-th impression, (in terms of CTR, conversion value etc.); $cost_i$ represents advertiser's cost for a winning the $i$-th impression; $click_i$ is an indicator variable referencing whether the $i$-th impression is clicked by user; $viewable_i$ is an indicator variable for whether the $i$-th impression is viewable to the user; $B$ is the total budget for a campaign; $C$ is the CPC performance goal; $V$ is the viewability goal.

To solve the problem, one needs to find the proper bidding strategy, which results in optimal $x_i$ that maximizes the target function under KPI constraints. In a second-price auction, the optimal bidding strategy takes the form of $v * m$, where $v$ is the expected impression value (e.g. $V * CTR$, where $V$ is the value of a clicked impression), and $m$ is a scaling parameter [17]. Bid price is determined according to the impression value and it increases as the value becomes larger.

To find the optimal bidding strategy, one could try to derive the static optimal $m$ from historical data. However, it is difficult—often prohibitively so—to derive static solutions from historical observations, as RTB environments are volatile and difficult to predict. Various factors contribute to this volatility—for example: bidders' strategies may change based on their expectations of how other bidders will act; seasonal changes are common for many markets; technological changes may compel strategic adjustments; and the traders and companies participating in markets will change over time.

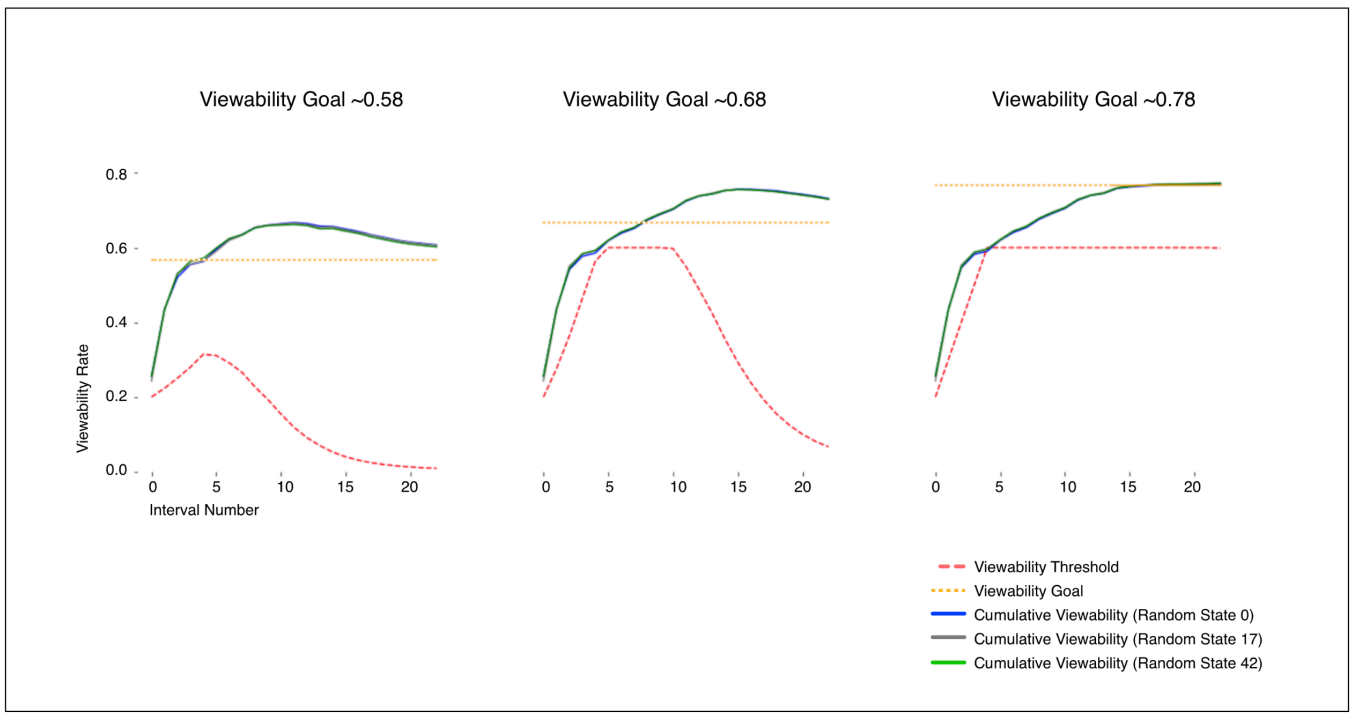

**Figure 3a: Controlling Viewability with the Viewability Threshold:**
Viewability threshold, viewability goal, and measured cumulative viewability from three tests in an RTB simulator (c.f. Section 6.1). Inventory was randomly selected from a dataset of previously won impressions according to the Random State value. We observe that the feedback control system is effective in adjusting the viewability threshold in response to the given goal—moving quickly at first, then more carefully, in order to reach the goal with minimal oscillation.

In order to address these challenges, we employ feedback control theory to generate a dynamic bidding strategy that can adapt to the non-stationary RTB environment in real time. The base bidding strategy still takes the form of $b = m * v$, but we make it dynamic by adjusting $m$ in real time through the feedback control method. Two additional levers, tolerance and viewability threshold, are introduced to improve the controllability and effectiveness of the system. A more detailed description of the levers is provided in Section 5.

## 3.2 Controller Module

The controller module takes observations about the current state of the system, compares this against a desired goal state, and generates a control signal output. This output is a single value that expresses how far the system is from its goal, in terms of a magnitude and direction; the actuator receives this value and moves levers accordingly.

In addition to considering the current state of the system, the control signal should encompass its rate of change and its prior history; therefore a PID module is used. The PID implementation includes three components [1], where KPI error is defined as $e_k(t) = GV_k(t) - KPI_k(t)$, and $GV$ is the goal KPI value.

- **Proportional**: The current *error value*—i.e., the difference between measured state and goal.
  - $P(k, t) = e_k(t)$
- **Integral**: The integral of the error value curve; this can be over the lifetime of operation or may be restricted to a shorter lookback window. Enables the system to account for long-term accumulated error. Also results in more predictable responses by the actuator.
  - $I(k, t) = \int_0^t e_k(t') \, dt'$
- **Derivative:** The derivative of the error value curve. Enables the actuator's response to account for trends in a quickly-changing measured state.
  - $D(k, t) = e_k(t) - e_k(t - 1)$

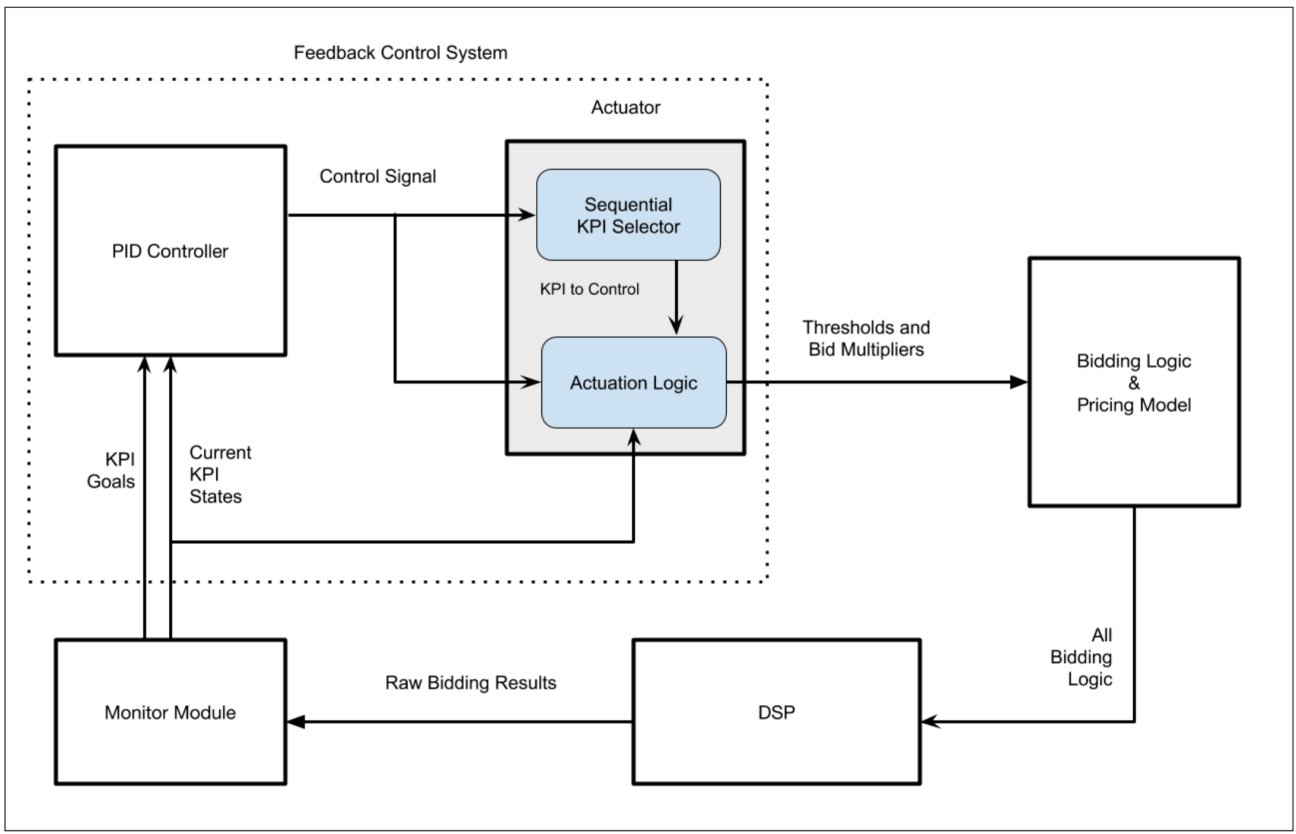

**Figure 3b: Feedback Control Loop System Diagram**

These three values will be combined into a single objective output: each term is multiplied by a gain coefficient (see Reproducibility section) in order to weight the terms by relative importance, then summed.

### 3.3 Actuator Module

Once the control signal is generated, it is received by the actuator. The actuator converts the raw control signal into a bid multiplier or threshold, according to the specific lever used to control the KPI in question, as discussed in Section 5. This is done with an exponentiation function, in which $m_{k,l}$ is the coefficient for KPI $k$ and lever $l$, and $w_{k,l}$ is the per-lever weight:

$$m_{k,l}(t+1) = w_{k,l}\, m_{k,l}(t)\, e^{\varphi k} \quad (1)$$

Constraints are placed on the final lever values, in order to avoid extreme bidding changes. For example, the viewability threshold must be between 0.01 and 0.6, and the adjustment to the threshold in a single interval cannot be of greater magnitude than ± 0.1.

The new bid multipliers and thresholds are used to generate new bidding logic, which is then uploaded to the DSP. The DSP bids for a preset period of time—in our case, 6 hours—and then our monitor module compiles the newest updates to the KPI states. This completes the feedback loop, allowing for a fast turnaround time to make changes and assess the resulting KPI changes.

## 4 Sequential Control

The basic challenge of improving multiple RTB goals is that an attempt to improve any one goal may affect the other goals, often in unpredictable ways. It follows that a straightforward way to address multiple goals in a feedback loop is to focus on a single goal at a time, allowing for a clear series of cause-and-effect events. (Our tests on a method that attempts to improve all KPIs simultaneously in each interval, as discussed in Section 6, shows notably worse performance than a one-at-a-time approach.) Therefore, we implemented two methods for sequential KPI control: Simple Sequential and Smart Sequential Control.

### 4.1 Simple Sequential

A straightforward approach to sequentially controlling multiple KPIs is to use a simple iterated method, as in Geyik, et al. [2016]; this is often referred to as the *Lexicographic method* [9, 5]. This approach requires the use of an objective ordering of all KPIs, such as in descending order of importance. In our case, this is decided according to the specific business needs of the client, so the process of deciding on a particular KPI order can be considered exogenous to this model.

This process simply starts at the highest priority KPI and determines whether it is acceptably close to its goal: if so, it moves on to the next KPI; if not, it stops the loop and chooses that KPI to control for the next interval.

In order to determine whether a KPI is sufficiently close to goal, we use the PID control signal. KPI $k$ is considered fully optimized if its control signal, $\varphi_k$, is zero; therefore, we have a threshold $t$, such that $k$ is considered acceptable if $abs(\varphi_k) \leq t$.

If all KPIs are considered acceptable, then no change will be made to any levers in that interval. If we find a KPI to be unacceptable, we will pass the choice of KPI to the actuator, which will update the lever value, as described in Section 3.3. As this is a sequential method, only one KPI will be passed to the actuator per time interval—typically every four to six hours.

### 4.2 Smart Sequential

Simple Sequential is effective at improving multiple KPIs—favoring higher-priority goals, while seeking improvements in lower-priority areas when possible. However, this method risks trading off small gains in the first priority KPI for large declines in the lower-priority KPIs. Furthermore, if the highest priority KPI cannot come close to being optimized, then all subsequent KPIs will be neglected as well [5]. Smart Sequential is an approach to rectify this.

Smart Sequential modifies the control signal for each KPI according to its place on the priority list, then chooses the KPI whose adjusted control signal has the greatest absolute value. The adjusted control signal, $\varphi'_k$, is generated in the following way, for KPI $k$ at priority $p$, with $K$ total KPIs:

$$\varphi'_k = B^{K-p}\, \varphi_k \quad (2)$$

B is the exponential base, with a default value of 2. Control signals are normalized, so the adjusted control signals are directly comparable between KPIs. This results in a mechanism which favors improving higher-priority KPIs, while still being able to switch to lower-priority KPIs when particularly in need of improvement. The value of B can be adjusted in our system at any time: a sufficiently high value will result in behavior similar to Simple Sequential, while setting B=1 will result in the system ignoring the priority order entirely. We find that setting B=2 provides a well-balanced medium between the extremes; this is the setting used in all tests in Section 6.

## 5 Levers and Adjustments

The final bid is a function of the predicted CTR; the exact shape of the bid function differs slightly between different revenue types. Here, without loss of generality, we base the discussion on the CPM revenue type. The bid function is shown in Figure 4a. The bid price is a linear function of CTR, capped by minimum bid and maximum bid; note that these specific values are predefined by our users on a case-by-case basis. This bid function can be altered by two levers, tolerance $t$, and bid multiplier $m$. In addition, viewability threshold $v$ is added to help control the viewability KPI. Therefore, there are in total three levers in our control system, and they are used to tune our bidding strategy to achieve the KPIs dynamically.

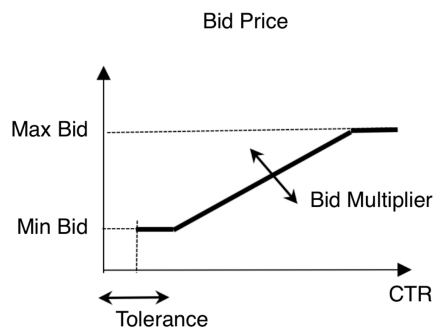

**Figure 4a: Bid price as a function of CTR. Tolerance and bid multiplier are the two levers that directly affect bid function.**

**Tolerance** enables our system to ignore individual auctions if the predicted CTR is too low. When tolerance is zero, we bid on everything; when tolerance is raised, any auctions for an impression with a predicted CTR below the tolerance level are ignored. This way, we concentrate our bidding on high-quality impressions that are relatively likely to be clicked. The tolerance lever has the potential to impact all three KPIs: it can reduce pacing because it raises selectivity of which impressions we bid on. It influences performance (e.g. CPA or CPC) by affecting the number of total clicks in the set of purchased impressions. In addition, it may affect average viewability, as viewability and engagement rates are often correlated [2].

**Bid multiplier** adjusts the slope of the linear bid function. In general, the bid price will be a linear function of predicted CTR, meaning bid price is higher for higher quality impressions. (Figure 4a) The bid multiplier has a potential impact on at least two KPIs: it influences performance because it directly changes the cost of media buying; and it affects pacing because the win rate of an impression directly depends on the bid price—the higher we bid, the more likely we are to win a given auction. The bid multiplier is likely to also affect viewability: naively, bid multiplier and average viewability rates should be directly correlated. However, real effects may vary, depending on the priorities of the other bidders.

**Viewability threshold** enables our system to ignore individual auctions if the predicted viewability is too low. Viewability predictions are provided to us by the DSP. If our viewability threshold is above zero, we will ignore any auction where the view probability is too low, or for which no probability is provided. This lever directly affects the average viewability of our purchased impressions. It also affects pacing in the same manner as tolerance: reducing our pool of available inventory can affect the ability to meet delivery goals. Finally, the viewability threshold is likely to affect performance: if other bidders are willing to raise their bids for inventory with a high probability of being viewed, then a higher viewability threshold will likely result in a raised CPA/CPC.

## 5.1 Table of Weights

From the description above, it is clear that each single lever influences multiple KPIs, and in turn each KPI can be controlled by multiple levers. It is challenging to decide which KPI to focus on, and which levers to pull at each time step. Therefore, when controlling a given KPI, all 3 levers can be pulled, and a matrix $W$ is used to connect levers and KPIs:

**Table 4b: Weights for Each Lever, When Controlling Each KPI**

| KPI to Control | Weight for Tolerance | Weight for Bid Multiplier | Weight for Viewability Threshold |
|---|---|---|---|
| Pacing | -0.5 | 0.5 | 0 |
| CPC | -0.5 | 0.5 | 0 |
| Viewability | 0 | 0 | 1 |

$W$ determines which levers should be used to control each KPI, and how much weight should be assigned to each lever. For example, if one wants to stop controlling pacing with viewability threshold lever, then $W_{PV}$ should be set to 0. The values in $W_{ij}$ can be values other than 0 or 1: they are simple multipliers, which can be assigned to levers based on their predicted effectiveness. In our current system, $W_{ij}$ are determined based on heuristic rules.

The sign of $W_{ij}$ indicates the direction to move a lever in order to raise a given KPI value. Therefore, because we understand that raising the viewability threshold will increase viewability, this weight/KPI combination should always be nonnegative if we intend to raise viewability. Conversely, if we understand that reducing tolerance will raise spend, the Pacing/Tolerance combination should never have a positive weight when we intend to increase pacing.

## 6 Experimental Results

This section covers experimental results for our PID feedback control approach, as well as for the methods for sequential multi-goal control. This is separated into tests conducted with an offline bid simulator, and a series of online tests against live traffic.

## 6.1 Offline Results

The offline experiment was conducted on a bidding simulator, using internal datasets representing previously purchased impressions. The first portion of this dataset is one week of impression records, to train our bid price predictor (the same method that is used in our production system). In order to generate inventory to bid on, we use a second week of impressions—which were all purchased subsequent to the first week—and randomly draw impressions from this set. The volume of this random selection is roughly equal to the average daily number of auctions into which we place a bid during the second week of data through the course of a single day. This becomes the source for a simulated day of impression auctions. We go in order, simulating a bid on each impression, as long as the impression meets all relevant criteria (e.g. sufficient probability of view, sufficient predicted CTR, etc.). Because our live auctions are second-price, we know the market-clearing price for all of our inventory. If the simulator places a bid, and it is equal to or greater than the market-clearing price, we count the auction as won. When all of the available inventory is either won or lost—or all of the daily budget is spent—the simulated bidding is ended, and we can find metrics on the performance of the won impressions.

The dataset for this test is comprised of 26 advertising campaigns, based on purchased impressions throughout the course of two separate weeks.

The first step of this test was to formulate a set of goals for each KPI. The goals are set as a function of the KPI values for the training week for each campaign, and they are set naively—they may not ever be attainable—and as such, they are useful for demonstrating the tradeoffs made by each optimization approach. The goals are:

- Raise spend by 50% over the training week
- Reduce CPA by 50% from the training week
- Raise average viewability by 20% vs. the training week

Because Simple Sequential and Smart Sequential require a KPI priority order, the following orders were tested:
- **Viewability, CPA, Pacing**
- **Pacing, Viewability, CPA**
- **CPA, Pacing, Viewability**

The following multi-goal methods were used:
- **Baseline:** Make no changes to any lever.
- **All At Once (AAO):** Attempt to improve all KPIs simultaneously—with no sequentiality or consideration of priority order.
- **Simple Sequential**
- **Smart Sequential**

For Smart Sequential, the amount by which higher-priority KPIs are weighted above those with lower priorities is adjustable; for these tests, we use the setting B=2, as discussed in Section 4.2 and Equation 2. With this setting, Smart Sequential does not strictly favor higher-priority KPIs— demonstrating the difference in approach versus Simple Sequential, which strongly favors higher priority goals. Note that the latter method can be closely approximated with Smart Sequential, by raising the exponential base in the multiplier.

All results show percent change relative to baseline.

**Table 6.1a: Viewability, CPA, Pacing**

| Method | Spend | CPA | Viewability |
|---|---|---|---|
| Smart | **-2.1%** | +22.3% | +20.5% |
| Simple | -30.5% | **-15.4%** | **+23.5%** |
| A.A.O. | -26.4% | +6.7% | +21.1% |

**Table 6.1b: Pacing, Viewability, CPA**

| Method | Spend | CPA | Viewability |
|---|---|---|---|
| Smart | +32.6% | +32.5% | +5.8% |
| Simple | **+40.1%** | +33.8% | +2.4% |
| A.A.O. | -21.3% | **+5.0%** | **+21.6%** |

**Table 6.1c: CPA, Pacing, Viewability**

| Method | Spend | CPA | Viewability |
|---|---|---|---|
| Smart | **+35.0%** | +33.6% | +4.5% |
| Simple | -55.1% | **-22.0%** | -3.1% |
| A.A.O. | -27.4% | +3.1% | **+21.1%** |

In the offline tests, Simple Sequential consistently yielded the greatest improvement on the first-priority KPI. Conversely, Smart Sequential is more indifferent to the KPI priority order, being likely to yield states in which the majority of KPIs are improved or remain satisfactory, at the expense of solely the first or second priority KPI.

The All At Once method performed worst on every KPI in Figure 6.1A; on Figure 6.1C, it only returned the best improvement for Viewability, the lowest-priority KPI. While it showed strong performance in Figure 6.1B for Viewability and CPA—the second and third KPIs, respectively—this came at the cost of a 22% decline in the first-priority, Spend. It follows that a sequential approach to feedback control is more effective in yielding improved outcomes than simply controlling all KPIs at all times.

## 6.2 Online Single-Goal Results

Single-goal tests were conducted to test feedback control on real traffic; because there was only one KPI to improve, sequential control was not used for these tests.

The objective of the single-goal online test was to improve cost per click (CPC) on a live RTB campaign. Almost 2 million impressions were purchased, across four test groups:
- **Control Group**
- **Feedback Control: Bid Multiplier (FC:BM)**
- **Feedback Control: Tolerance (FC:T)**
- **Feedback Control: Multilever (FC:M)** Bid Multiplier and Tolerance, as discussed in Section 5

The campaign was run continuously for 17 days, with a targeting scheme to ensure that the test groups would not bid against each other for the same impressions.

All results show percent change relative to control.

**Table 6.2a: Feedback Control for CPC**

| Method | Purchased Impression Volume | CPC |
|---|---|---|
| FC:BM | -0.27% | **-68.7%** |
| FC:T | -0.30% | -5.2% |
| FC:M | -0.30% | -62.6% |

In these tests, feedback control with the bid multiplier lever yielded in a significant reduction in CPC, with no meaningful decrease in purchase volume. The tolerance lever resulted in only a minor improvement against control, suggesting that low-CTR inventory could be purchased at a commensurately low price. The combined multilever approach yielded a significant improvement, but less than the bid multiplier lever alone.

## 6.3 Online Multi-Goal Results

A multi-goal test was conducted on real traffic, used to assess sequential feedback control. The KPI priority order was Pacing, Viewability; Simple Sequential control was used because spending the entire budget is a crucial goal in this campaign. The tests cover four groups:
- **Control group**
- **Default Multilever setup** (Table 4b)
- **Multilever Custom 1**, which adjusts the view threshold weights for the Pacing KPI to 0.1.
- **Multilever Custom 2**, which adjusts the view threshold weights for the Pacing KPI to 0.3.

Approximately 1.3 million impressions were purchased for this test; all results show percent change relative to control.

**Table 6.3a: Feedback Control for Pacing, Viewability**

| Method | Purchased Impression Volume | Viewability | CPC |
|---|---|---|---|
| Multilever: Default | +52.9% | -0.87% | -18.4% |
| Multilever: Custom 1 | +48.4% | **-0.51%** | **-25.7%** |
| Multilever: Custom 2 | **+57.2%** | -1.53% | -7.2% |

Here we see that all Multilever solutions significantly improved the first KPI—pacing—while holding viewability largely steady. This was done without purchasing low-quality inventory, as shown by the reduction in CPC against baseline.

## 7 Conclusion

In this paper, we demonstrated a method for using a feedback loop with a PID module to control and improve media buying KPIs. We provided a method to associate multiple levers with each KPI, and to adjust the magnitude by which each lever is used. We presented two methods for balancing goals for an arbitrary number of KPIs: Simple Sequential and Smart Sequential. Finally, we showed that these methods perform better than baseline, and additionally perform better than trying to improve all KPIs at once.

For further work, we would propose a multi-arm bandit solution to optimizing the various configuration options: Simple versus Smart Sequential; the magnitude of weights for each lever; and the exponential base for Smart Sequential. Furthermore, we would investigate using reinforcement learning, rather than the PID output and exponentiation model, to decide how to adjust each lever for the chosen KPI. This method could yield significant improvements over an iterative approach to adjusting KPI levers, and would be an exciting research direction.

## ACKNOWLEDGMENTS
The authors would like to thank Adam Cushner, Jake Grabczewski, Victor Seet, and Tobias Sutters for their contributions.

## REFERENCES
[1] Karl Johan Åström and Richard M. Murray. 2008. Feedback Systems: An Introduction for Scientists and Engineers. Princeton University Press, Princeton, NJ.
[2] David Bounie, Morrisson Valérie, and Martin Quinn. 2017. Do You See What I See? Ad Viewability and the Economics of Online Advertising. *SSRN Electronic Journal*. DOI: https://doi.org/10.2139/ssrn.2854265
[3] Andreas Doerr, Duy Nguyen-Tuong, Alonso Marco, Stefan Schaal, and Sebastian Trimpe, 2017. Model-based policy search for automatic tuning of multivariate PID controllers. In *Proc. 2017 IEEE International Conference on Robotics and Automation,* (May 2017), pp5295-5301. DOI: https://doi.org/10.1109/ICRA.2017.7989622
[4] Sahin Cem Geyik, Sergey Faleev, Jianqiang Shen, Sean O'Donnell, and Santanu Kolay. 2016. Joint Optimization of Multiple Performance Metrics in Online Video Advertising. In *Proceedings of the 22nd ACM SIGKDD International Conference on Knowledge Discovery and Data Mining* (KDD '16). Association for Computing Machinery, New York, NY, USA, 471–480. DOI:https://doi.org/10.1145/2939672.2939724

[5] Brendan Kitts, Michael Krishnan, Ishadutta Yadav, Yongbo Zeng, Garrett Badeau, Andrew Potter, Sergey Tolkachov, Ethan Thornburg, and Satyanarayana Reddy Janga. 2017. Ad Serving with Multiple KPIs. In *Proceedings of the 23rd ACM SIGKDD International Conference on Knowledge Discovery and Data Mining* (KDD '17). Association for Computing Machinery, New York, NY, USA, 1853–1861. DOI:https://doi.org/10.1145/3097983.3098085

[6] Kuang-Chih Lee, Ali Jalali, and Ali Dasdan. 2013. Real time bid optimization with smooth budget delivery in online advertising. In *Proceedings of the Seventh International Workshop on Data Mining for Online Advertising* (ADKDD '13). Association for Computing Machinery, New York, NY, USA, Article 1, 1–9. DOI:https://doi.org/10.1145/2501040.2501979

[7] H. Brendan McMahan, Gary Holt, D. Sculley, Michael Young, Dietmar Ebner, Julian Grady, Lan Nie, Todd Phillips, Eugene Davydov, Daniel Golovin, Sharat Chikkerur, Dan Liu, Martin Wattenberg, Arnar Mar Hrafnkelsson, Tom Boulos, and Jeremy Kubica. 2013. Ad click prediction: a view from the trenches. In *Proceedings of the 19th ACM SIGKDD International Conference on Knowledge Discovery and Data Mining* (KDD '13). Association for Computing Machinery, New York, NY, USA, 1222–1230. DOI:https://doi.org/10.1145/2487575.2488200

[8] PID Tuning Guide: A Best-Practices Approach (2010). Retrieved from https://controlstation.com/cstuner-pid-tuning-guide/

[9] Ivan P. Stanimirovic. 2012. Compendious lexicographic method for multi-objective optimization. *Series Mathematics Informatics* 27 (2012), 55-66.

[10] Jun Wang, Weinan Zhang, and Shuai Yuan. 2016. Display Advertising with Real-Time Bidding (RTB) and Behavioural Targeting. arXiv:1610.03013. Retrieved from https://arxiv.org/abs/1610.03013.

[11] Jian Xu, Kuang-chih Lee, Wentong Li, Hang Qi, and Quan Lu. 2015. Smart Pacing for Effective Online Ad Campaign Optimization. In *Proceedings of the 21th ACM SIGKDD International Conference on Knowledge Discovery and Data Mining* (KDD '15). Association for Computing Machinery, New York, NY, USA, 2217–2226. DOI:https://doi.org/10.1145/2783258.2788615

[12] Xun Yang, Yasong Li, Hao Wang, Di Wu, Qing Tan, Jian Xu, and Kun Gai. 2019. Bid Optimization by Multivariable Control in Display Advertising. In *Proceedings of the 25th ACM SIGKDD International Conference on Knowledge Discovery & Data Mining* (KDD '19). Association for Computing Machinery, New York, NY, USA, 1966–1974. DOI:https://doi.org/10.1145/3292500.3330681

[13] Shuai Yuan, Jun Wang, and Xiaoxue Zhao. 2013. Real-time bidding for online advertising: measurement and analysis. In *Proceedings of the Seventh International Workshop on Data Mining for Online Advertising* (ADKDD '13). Association for Computing Machinery, New York, NY, USA, Article 3, 1–8. DOI:https://doi.org/10.1145/2501040.2501980

[14] Weinan Zhang, Ye Pan, Tianxiong Zhou, and Jun Wang. 2015. An Empirical Study on Display Ad Impression Viewability Measurements. arXiv: 1505.05788v1. Retrieved from https://arxiv.org/pdf/1505.05788

[15] Weinan Zhang, Yifei Rong, Jun Wang, Tianchi Zhu, and Xiaofan Wang. 2016. Feedback Control of Real-Time Display Advertising. In *Proceedings of the Ninth ACM International Conference on Web Search and Data Mining* (WSDM '16). Association for Computing Machinery, New York, NY, USA, 407–416. DOI:https://doi.org/10.1145/2835776.2835843

[16] Weinan Zhang, Shuai Yuan, and Jun Wang. 2014. Optimal real-time bidding for display advertising. In *Proceedings of the 20th ACM SIGKDD international conference on Knowledge discovery and data mining* (KDD '14). Association for Computing Machinery, New York, NY, USA, 1077–1086. DOI:https://doi.org/10.1145/2623330.2623633

[17] Weinan Zhang, Tianxiong Zhou, Jun Wang, and Jian Xu. 2016. Bid-aware Gradient Descent for Unbiased Learning with Censored Data in Display Advertising. In *Proceedings of the 22nd ACM SIGKDD International Conference on Knowledge Discovery and Data Mining* (KDD '16). Association for Computing Machinery, New York, NY, USA, 665–674. DOI:https://doi.org/10.1145/2939672.2939713

[18] Wen-Yuan Zhu, Chun-Hao Wang, Wen-Yueh Shih, Wen-Chih Peng, Jiun-Long Huang. 2017. SEM: A Softmax-based Ensemble Model for CTR estimation in Real-Time Bidding advertising. *2017 IEEE International Conference on Big Data and Smart Computing* (2017), 5-12. DOI:10.1109/BIGCOMP.2017.7881698

# Appendix: Reproducibility

## Gain Terms

Our PID implementation contains a default gain multiplier for each PID term:

- **Proportional:** $K_P^k = \frac{1}{GV_k(t)}$. Purpose is to normalize the Proportional term so that control signals from different KPIs may be compared. This is included in the gain coefficients for the other two terms as well, thereby normalizing the entire PID control signal.
- **Integral:** $K_I^k = \frac{K_P^k}{T_I}$, where $T_I$ is the number of intervals being used to compute the integral term. By default, $T_I = 10$.
- **Derivative:** $K_D^k = T_D K_P^k$, where $T_D$ is the number of intervals being used to compute the derivative term. By default, $T_D = 2$.

These default gain terms can be overridden, but have not been adjusted from default in practice.

## Lever Constraints

Each lever has a fixed minimum and maximum value, and a maximum change in absolute value per interval.

| Lever | Minimum Value | Maximum Value | Maximum Change Per Interval |
|---|---|---|---|
| Viewability Threshold | 0.01 | 0.6 | 0.1 |
| Bid Multiplier | 0.1 | 10.0 | 1.0 |
| Tolerance | 0 | 95th percentile of the predicted CTR | 1/5th of the 50th percentile of predicted CTR |